\documentclass[a4paper]{jpconf}
\usepackage{graphicx}
\renewcommand{\r}{{\bf r}}
\newcommand{\rv}{\vec{\rm r}}

\newcommand{\be}{\begin{eqnarray}}
\newcommand{\ee}{\end{eqnarray}}

\begin{document}
\title{
Pairing correlations in exotic nuclei} 

\author{ H. Sagawa$^{1,2}$  and K. Hagino$^3$}
\address{$^1$Center for Mathematics and Physics, University of Aizu,
Aizu-Wakamatsu, 965-8580 Fukushima, Japan}

\address{$^2$Nishina Center, RIKEN, Wako,  351-0198, Japan}
\address{$^3$Department of Physics, Tohoku University, Sendai, 
 980-8578, Japan}
\ead{sagawa@u-aizu.ac.jp}

\begin{abstract}
The BCS and HFB theories which can accommodate the pairing correlations in the ground states of 
atomic nuclei are presented. As an application of the pairing  theories , 
we investigate the spatial extension of weakly bound Ne and C isotopes 
by taking into account the pairing correlation with 
the Hartree-Fock-Bogoliubov (HFB) method 
and a 3-body model, respectively. 
We show that the odd-even staggering in the reaction cross sections 
of $^{30,31,32}$Ne and $^{14,15,16}$C 
are successfully reproduced, and thus the staggering can be attributed 
to the unique role of pairing correlations  in  nuclei far from the stability line.  
A correlation  between a one-neutron separation energy 
and the anti-halo effect
is demonstrated for 
$s$- and $p$-waves  
using the HFB wave functions.
We  also  propose  effective density-dependent
pairing interactions which reproduce both the neutron-neutron ($nn$) 
scattering length at zero density and the neutron pairing gap in
uniform matter.  Then, we apply these interactions to study  pairing gaps in semi-magic finite 
 nuclei, such as Ca, Ni, Sn and Pb isotopic chains.
\end{abstract}

\section{Introduction} 
It has been known that the pairing correlations
 play an important role in finite and also infinite 
nuclear systems. 
Just after the BCS theory was proposed as a fundamental theory of metallic superconductor, 
 Bohr, Mottelson and Pines 
proposed a possible analogy of superfluidity in atomic nuclei in 1958\cite{BMP58}.  
The most prominent evidence  for  pairing correlation in nuclei
is found
in the
odd-even staggering in binding energies and the  energy  gap in the excitation
 spectrum of
even-even nuclei 
in contrast to  a  compressed quasi-particle spectrum
in
odd-A nuclei \cite{BMP58,BM69,BB05}.
There are also dynamical effects of pairing
correlations seen in the moment of inertia associated with nuclear rotation and
large amplitude collective motion
\cite{BB05,Bertsch12,Matsuyanagi12}.
The Hartree-Fock (HF)+BCS method and  Hartree-Fock-Bogoliubov (HFB) method
have been commonly used to study the ground state properties of superfluid
nuclei
in a broad mass region \cite{Ben03,Stoi06,dug01,mar07}.

As new phenomena in nuclei near the neutron-drip line,  
large odd-even staggering (OES) phenomena have been 
revealed  experimentally  in reaction cross sections 
%
 of  the isotopes $^{14,15,16}$C \cite{FYZ04}, $^{18,19,20}$C \cite{Ozawa00}, 
$^{28,29,30}$Ne\cite{Takechi10}, $^{30,31,32}$Ne \cite{Takechi10}, and $^{36,37,38}$Mg \cite{Takechi11}.
In Ref. \cite{HS11}, we have argued that 
the pairing correlations play an essential role 
in these OES.  
That is, the OES in reaction cross sections 
is intimately related to 
the so called pairing anti-halo effect discussed in Ref. \cite{Benn00}.
In this lecture , 
we  discuss the pairing correlations close to the zero energy
by a  Hartree-Fock Bogoliubov (HFB) method. 
This problem is also related with the superfluidity 
of neutron gases in the outer crust of neutron 
stars \cite{Schuck11}.

 Recently, we proposed new types of density dependent contact pairing
interaction, which reproduce pairing gaps in  a wide range of 
nuclear mass table \cite{mar07}.  
We discussed  also the relation between the proposed paring interactions and 
the pairing gaps in symmetric and neutron
matters obtained by 
a microscopic treatment based on the nucleon-nucleon interaction.
We will show  the necessity of the isovector type pairing 
interaction  on top of the isoscalar term to reproduce 
systematically nuclear empirical pairing gaps.

This lecture note is organized as follows.  Section 2 gives basic formulas of BCS and HFB theories. 
In section 3, we will discuss the odd-even staggering in the reaction cross sections in nuclei 
near the neutron drip line.  Section 4 is devoted to study the isospin dependent pairing interaction 
for the study of pairing correlations in semi-magic nuclei.  

\section{BCS and  HFB theories}
A very important generalization of the HF theory is to accommodate pairing interaction on  top of
the mean field.  The general theoretical framework with the pairing correlations for
 the single-particle orbitals is 
called Hartree-Fock-Bogoliubov  (HFB) equations,  that is also  
called the Bogoliubov-deGenne equations in condensed 
matter physics.  A simpler version  of HFB  
is called the BCS theory which has been   often employed  
in nuclear physics.  In dealing with even-even nuclei, the HF equations are invariant under time 
reversal.  This implies that each orbital $i$ has its time-reversed partner $\bar{i}$, and the 
two orbits have the same single-particle energy $\varepsilon _i$.  The basic BCS  Ansatz  for the 
many-particle wave function is
\be \label{BCS}
|\mbox{BCS}\rangle=\prod_{i>o}(u_i+v_ia^{\dagger}_ia^{\dagger}_{\bar{i}})|0\rangle
\ee
where $a^{\dagger}_i$ is the particle creation operator acting on the HF vacuum $|0\rangle$. 
 The parameters $u_i$ and $v_i$  will be determined by minimizing
 the expectation  value  of the Hamiltonian.   The  normalization of the state requires
\be
 |u_i|^2+|v_i|^2=1.
\ee
The BCS state (\ref{BCS}) is further rewritten to be
\be
|\mbox{BCS}\rangle
  \propto  \mbox{exp}\biggl( \sum_{i>0}\frac{v_i}{u_i}a^{\dagger}_ia^{\dagger}_{\bar{i}} \biggr)|0\rangle.
\ee
The BCS state is not an eigenstate of the particle number.
In condensed matter physics, this  does not cause  
any serious problem since the number of particles  is  
close to  the Avogadro number $\sim$10$^{23}$.  On the other hand, in nuclear physics, 
the number of nucleons is  in the  order of at most 200 so that we have to take care of the number conservation in some way. One possible way is to introduce a Lagrange multiplier term in the Hamiltonian
\be \label{conH}
H'=H-\lambda\hat{N}.
\ee
Then the particle number expectation value can be fixed to the desired value $N$
\be
\langle\mbox{BCS}|\hat{N}|\mbox{BCS}\rangle=2\sum_{i>0}v_i^2=N
\ee
 on  average.  The quasi-particles are introduced by a  unitary  transformation, so called 
 the  Bogoliubov-Valatin transformation
\be \label{BVT}
\alpha^{\dagger}_i=u_{i}a_i^{\dagger}-v_{i}a_{\bar{i}} \nonumber \\
\alpha_{\bar{i}}=u_{i}a_{\bar{i}}+v_{i}a_i^{\dagger}  \label{qpt}
\ee
where 
$a_i$ is
the physical (or bare) annihilation operator.
For the unitary  transformation, the quasi-particles preserve the anti-commutation relations
\be
\{\alpha_i,\alpha^{\dagger}_j\}=\delta_{i,j}, \,\,\,\,\,\,
 \{\alpha_i,\alpha_j\}=\{\alpha^{\dagger}_i,\alpha^{\dagger}_j\}=0.
\ee
The BCS vacuum is defined by an equation
\be 
\label{BCS-a}
\alpha_i|\mbox{BCS}\rangle=0.  
\ee
We can prove that the definition  (\ref{BCS-a}) is equivalent to the BCS vacuum (\ref{BCS}).
Firstly,  the definition (\ref{BCS-a})  can read
\be \label{BCS-1}
|\mbox{BCS}\rangle=\prod_{i\in all}\alpha_i |0\rangle =\prod_{i>0}\alpha_i\alpha_{\bar{i}}|0\rangle .
 \ee
By performing the transformation (\ref{qpt}) from  the quasi-particles to the bare particles in Eq. 
(\ref{BCS-1}), we obtain
\be
|\mbox{BCS}\rangle&=&\prod_{i>0}\{u_{i}^2a_ia_{\bar{i}}+u_{i}v_{i}(1-a_i^{\dagger}a_i) -
u_{i}v_{i}a_{\bar{i}}^{\dagger}a_{\bar{i}}+v_{i}^2a_i^{\dagger}a_{\bar{i}}^{\dagger}\}|0\rangle \nonumber \\
 &=&\prod_{i>0}v_i\{u_i+v_ia_i^{\dagger}a_{\bar{i}}^{\dagger}\}|0\rangle \propto \prod_{i>0}\{u_i+v_ia_i^{\dagger}a_{\bar{i}}^{\dagger}\}|0\rangle .  
\ee
The hamiltonian density can be evaluated for the BCS state as
\be
\langle \mbox{BCS}|H'|\mbox{BCS} \rangle =\sum_{i}(\varepsilon_i-\lambda)v_i^2-\sum_{i,j>0}V_{i\bar{i}j\bar{j}}u_iv_iu_jv_j
\ee
The variation of the Hamiltonian density with respect to the $v_i$ (or  equivalently  $u_i$ ) gives 
a BCS gap equation to be solved;
\be
\Delta_i=\sum_{j>0}\frac{2j_j+1}{2}V_{i\bar{i}j\bar{j}}u_jv_j
\ee
with
\be
u_i^2=\frac{1}{2}\biggl( 1+\frac{\varepsilon_i-\lambda}{\sqrt{(\varepsilon_i-\lambda)^2+\Delta_i^2}}   \biggr)   \nonumber \\
v_i^2=\frac{1}{2}\biggl( 1-\frac{\varepsilon_i-\lambda}{\sqrt{(\varepsilon_i-\lambda)^2+\Delta_i^2}}
                         \biggr)
\ee
The quasi-particle energy $E_i$ is evaluated by a commutator relation
\be
[H', \alpha^{\dagger}_i]=E_i
\ee
to be
\be
 E_i=\sqrt{(\varepsilon_i-\lambda)^2+\Delta_i^2}.
\ee

The BCS theory is generalized to HFB theory by using the following  unitary  transformation.  
The  Hartree-Fock-Bogoliubov (HFB) transformation for quasi-particle can read
\be
\alpha^{\dagger}_i=\sum_iu_{ij}a_j^{\dagger}-v_{ij}a_{\bar{j}} \nonumber \\
\alpha_{\bar{i}}=\sum_iu_{ij}a_{\bar{j}}+v_{ij}a_j^{\dagger}  
\ee
where 
 $u_{ij}$ and $v_{ij}$ are the upper and the lower components of BCS transformation, 
respectively. The normalization condition for the $u_{ij}$ and $v_{ij}$ factors is given by
\be
\sum_ju_{ij}^2+v_{ij}^2=1.
\ee
The HFB model can be generalized  to  the coordinate space representation  as \cite{B00}
\be
\alpha^{\dagger}_i=\int d\r\{u_i(\r)\psi^{\dagger}(\r)+v_i^*(\r)\psi(\r)\} \nonumber \\
\alpha_i=\int d\r\{u_i^*(\r)\psi(\r)+v_i(\r)\psi^{\dagger}(\r) \}
\ee
where $\psi^{\dagger}(\r)$ and $\psi(\r)$ are the  creation and the annihilation operators of bare particle in the coordinate space representation and obey the anti-commutator relation
\be
\{\psi(\r), \psi^{\dagger}(\r')\}=\delta(\r-\r').
\ee
The orthonormal conditions of $u_i(\r)$ and $v_i(\r)$ functions  are given by
\be
\int d\r \{u_i^*(\r)u_j(\r)+v_i^*(\r)v_j(\r)\}=\delta_{i,j}  \nonumber \\
\sum_i\{u_i^*(\r)u_i(\r')+v_i^*(\r)v_i(\r')\}=\delta(\r-\r').
\ee
The density and the pair density (abnormal density) are obtained as 
\be
\rho(\r,  \r')=\langle\mbox{HFB}|\psi^{\dagger}(\r)\psi(\r')|\mbox{HFB}\rangle=\sum_iv_i(\r)v_i^*(\r')\\
\kappa( \r, \r')=\langle\mbox{HFB}|\psi(\r)\psi(\r')|\mbox{HFB}\rangle=\sum_iv_i(\r)u_i(\r') 
\ee
where the HFB vacuum is defined by
\be
\alpha_i|\mbox{HFB}\rangle=0.
\ee
The hamiltonian density for the constrained Hamiltonian $H'$ (\ref{conH}) is expressed to be
\be \label{HD}
\langle\mbox{HFB}|H'|\mbox{HFB}\rangle=Tr\{(T-\lambda)\rho\}+\frac{1}{2}Tr\{V\rho\rho\}+\frac{1}{2}\{V\kappa\kappa\}
\ee
where $T$ and $V$ are the kinetic density and the hamiltonian density due to  a  two-body interaction, 
respectively.

The HFB equations are obtained by the variation of the hamiltonian density in Eq. (\ref{HD}) 
 with respect to the  $u_i(\r)$ and $v_i(\r)$ functions to be
\be \label{HFB}
\left( \begin{array}{cc}
T+V_{HF}-\lambda  & \Delta(\r)  \\
\Delta(\r) & -T -V_{HF}+\lambda \end{array}
 \right) \left(  \begin{array}{c} 
                      u_i(\r) \\    v_i(\r)   \end{array} \right)
 = E_i\left(  \begin{array}{c} 
                      u_i(\r) \\    v_i(\r)   \end{array} \right). 
\ee
where $V_{HF}$ is the HF potential,
\be
V_{HF}=Tr\{V\rho\}
\ee
and $\Delta(\r)$ is the pairing gap potential,
\be
\Delta(\r)=Tr\{V\kappa\},
\ee
respectively.  
 In Eq. (\ref{HFB}),  $u_i(\r)$ and  $v_i(\r)$ are radial wave functions and  
could have  different shapes from  the HF single particle wave function, which is the 
eigenstate of HF hamiltonian $h_{HF}=T -V_{HF}$.   In contrast,  in the BCS model, 
 the $u_i$ and $v_i$ are factors and 
 simply multiplied to the  HF single particle wave function $\varphi_i$ 
by the Bogoliubov-Valatin  transformation (\ref{BVT}).

A question is whether the HFB is necessary for realistic  framework  
to treat the pairing correlations in
 nuclei.  When the active orbitals are well bound as in nuclei along the valley of stability, 
 the single-particle wave functions are not influenced  by the pairing field and the BCS model could 
be entirely adequate.  This might not be the case in nuclei near the drip lines, where the
 pairing correlations would make the difference between bound and unbound orbitals. 
In a recent paper, we showed that  a  major part of  the effects of HFB 
 could be included by a perturbative 
treatment of the BCS wave functions even in nuclei far from the valley of stability \cite{HS-pert}.  

\section{Halo and Pairing correlations}
  The pairing correlations play a crucial role to develop the halo structure of loosely bound nuclei.
At the same time, the pairing will act to prevent the divergence of very loosely bound nucleons in
the mean field.  Let us discuss how the single-particle wave function will expand its tail in the 
mean field potential and how the pairing correlations will prevent the exponential growth of 
expansion to be infinity.  In the mean field model, the single particle wave function
\be
 \psi_{lm}(\r)=\varphi_{lm}(r)Y_{lm}(\hat{r})=\frac{u_l(r)}{r}Y_{lm}(\hat{r})
\ee
is calculated with a Schr\"odinger equation
\be
 \biggl( -\frac{\hbar^2}{2m}\frac{d^2}{dr^2}+\frac{l(l+1)\hbar^2}{2mr^2 }+V_{HF}(r)-
 \varepsilon _{l} \biggr) u_l(r)=0 .
\ee
The asymptotic behavior
 in the limit $r \rightarrow \infty$ with $V_{HF} \rightarrow 0$ 
is given by
the modified spherical Bessel function and behaves
\be
u_{l=0}(r) \rightarrow \exp(-\kappa  r) \,\,\,\,\mbox{in the limit}  \,\,\,r \rightarrow \infty
\ee
for the angular momentum $l=0$ case.  Then the mean square radius for $s-$wave is evaluated to be
\be
\langle r^2\rangle _{l=0}=\frac{\int r^2u_{l=0}(r)^2dr}{\int u_{l=0}(r)^2dr} \propto 
 \frac{1}{\kappa^2}=\frac{\hbar^2}{2m|\varepsilon|} \rightarrow \infty
\ee 
which diverges in the limit of very loosely bound case $|\varepsilon| \rightarrow 0$.  
 It is also shown in the case of  $p-$wave case, $l=1$,  that the mean
 square 
radius also diverges  but  slightly slowly  proportional to $1/\sqrt{|\varepsilon|}$ \cite{Riisager}.  On the other hand, 
the wave function with $l\ge2$ does not show any divergence of mean square radius in the limit of 
$|\varepsilon|  \rightarrow 0$ \cite{Sagawa92}. 

In contrast, the upper component of a HFB wave function, which is 
relevant to the density distribution, behaves \cite{doba} 
\begin{equation}
  v(r) \propto \exp(-\beta r), 
\end{equation}
where $\beta$ is proportional to the square root of 
 the quasi-particle energy $E$,
\begin{equation}
\beta =\sqrt{\frac{2m}{\hbar ^2}(E-\lambda)},  
\end{equation}
$\lambda$ being the chemical potential. 
If we evaluate the quasi-particle energy 
in the BCS approximation 
or HFB with canonical basis, it is given as 
\begin{equation}
E=\sqrt{(\varepsilon -\lambda)^2 +\Delta^2}, 
\end{equation}
where 
$\Delta$ is the pairing gap.
For a weakly bound single-particle state with 
 $\varepsilon \sim 0$ and $\lambda  \sim 0$, the asymptotic 
behavior of the wave function 
$ v(r)$ is therefore determined by the gap parameter as, 
\begin{equation}
 v(r)\propto  \exp\left(-\sqrt{\frac{2m}{\hbar^2}\Delta}\cdot r\right).
\label{HFB-r}
\end{equation}
The radius of the HFB wave function will then 
be given in the limit of small separation energy $|\varepsilon|
 \rightarrow 0$ as 
\begin{equation}
 \langle r^2\rangle_{\rm \mbox{HFB}}
=\frac{\int r^2 |v(r)|^2 d{\bf r}}{\int  |v(r)|^2 d{\bf r}} \propto 
   \frac{1}{\beta^2} \rightarrow \frac{\hbar^2}{2m\Delta}.  
\end{equation}
As we show, the gap parameter  $\Delta$ stays finite even in the zero energy 
 limit of $\varepsilon$ with a density dependent pairing interaction.
Thus the extremely large extension of a halo wave function 
in the HF field will be reduced substantially by the pairing field 
and the root-mean-square (rms) 
radius of the HFB wave function will not diverge. 
This is called the 
anti-halo effect due to the pairing correlations \cite{Benn00}.

\begin{figure}
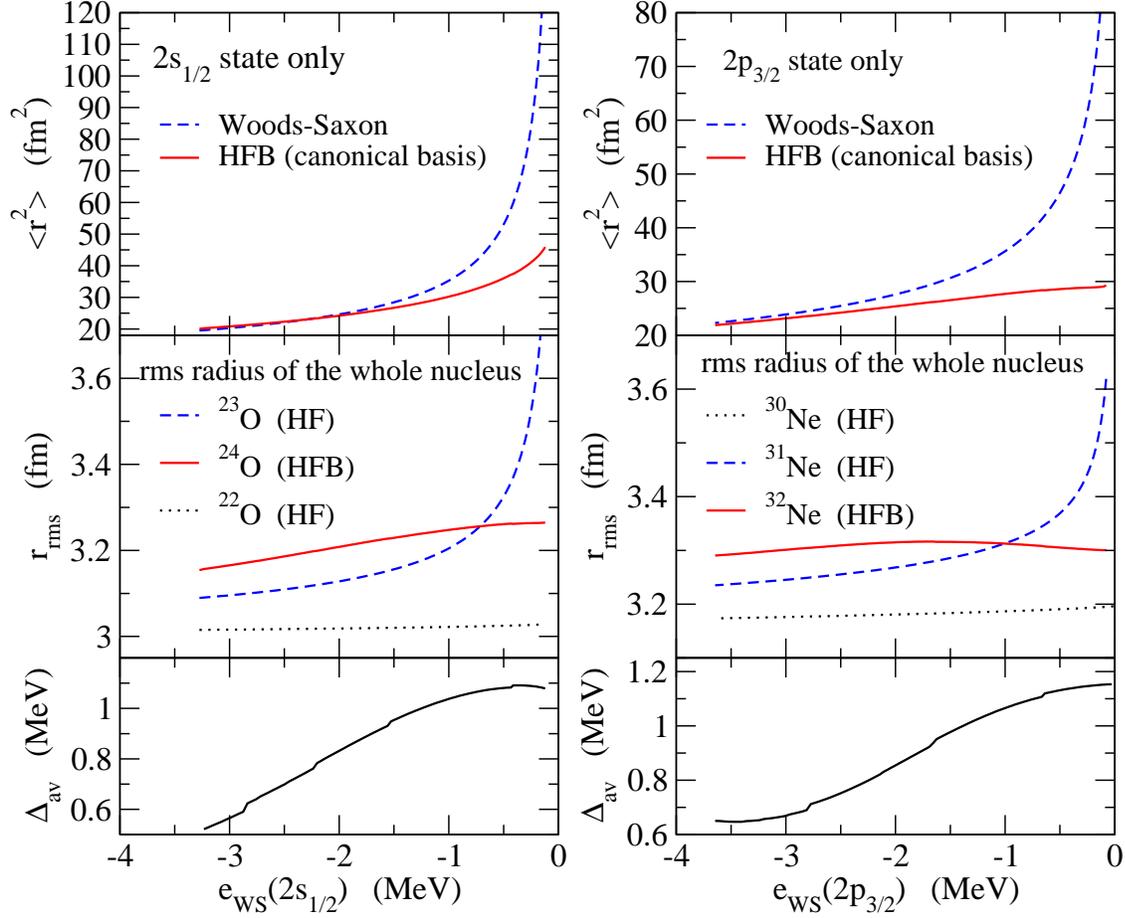
 
\includegraphics[scale=0.6,clip]{fig1}
\includegraphics[scale=0.6,clip]{fig2}
\caption{(Color online) 
(left panel) The mean square radii and the average paring gap as a function of 
the single particle energy $\varepsilon_{\rm WS}$ in a Woods-Saxon mean-field 
potential. 
The top panel shows the mean square radius of the 2s$_{1/2}$ wave function 
with and without the pairing correlation, denoted by 
HFB and Woods-Saxon, respectively.
The middle panel shows the rms radii 
for $^{22}$O (the dotted line), $^{23}$O (the dashed line), 
and $^{24}$O (the solid line), obtained with the Hartree-Fock ($^{22}$O and 
$^{23}$O) and the Hartree-Fock-Bogoliubov ($^{24}$O) calculations. 
The bottom panel shows the 
results of the HFB calculations for the 
average paring gap of $^{24}$O. (right panel) Same as the left panel, but for the 2$p_{3/2}$ state and for 
$^{30,31,32}$Ne isotopes. \label{Ne}
}
\end{figure}

We now numerically carry out 
mean-field calculations with a Woods-Saxon (WS) potential and also 
HFB calculations using the single-particle wave functions in the WS potential. 
Most of following materials of this section are taken from refs. \cite{HS11}.  
As examples of $s$-wave and $p$-wave states, we choose the 2$s_{1/2}$ state 
in $^{23}$O and 2$p_{3/2}$ state in $^{31}$Ne, respectively. 
Although $^{31}$Ne is most likely a deformed nucleus\cite{H10,UHS11}, for 
simplicity we assume a spherical Woods-Saxon mean-field potential. 
Notice that a Woods-Saxon potential with a 
large diffuseness parameter $a$ yields the 2$p_{3/2}$ state which is 
lower in energy than the 
1$f_{7/2}$ state, as was shown in Ref. \cite{HSCB10}. We use a similar 
potential with $a$=0.75 fm 
as in Ref. \cite{HSCB10} for $^{31}$Ne, while that in Ref. 
\cite{HS05} for $^{23}$O. 
For the HFB calculations, we use the density-dependent contact pairing 
interaction of surface type, in which the parameters are adjusted in 
order to reproduce the empirical neutron pairing gap for $^{30}$Ne \cite{YG04}. 
While we fix the Woods-Saxon potential for the mean-field part, the pairing 
potential is obtained self-consistently with the contact interaction. 

The left top panel of Fig. \ref{Ne}  shows the mean square radius 
of the 2$s_{1/2}$ state for $^{23}$O, 
while right top panel  of Fig. \ref{Ne} shows the mean square radius of the 
2$p_{3/2}$ state for $^{31}$Ne. 
In order to investigate the dependence on 
the single-particle energy, we vary the depth of the Woods-Saxon wells 
for the $s_{1/2}$ and $p_{3/2}$ states for $^{23}$O and $^{31}$Ne, respectively. 
The dashed lines are obtained with the single-particle wave functions, while 
the solid lines are obtained with the wave function for the canonical 
basis in the HFB calculations. 
One can see an extremely large increase of the radius 
of the WS wave function for both the $s$-wave and $p$-wave states 
in the limit of $\epsilon_{\rm WS}\rightarrow 0$. In contrast, 
the HFB wave functions show only a small increase of radius 
even in the limit of $\epsilon_{WS}\rightarrow 0$.  
This feature remains the same even when 
the contribution of the other orbits are taken into account, as shown 
in the middle panel of Figs. \ref{Ne}. 
Due to the pairing effect in the continuum,
the HFB calculations yield a larger radius than the HF calculations 
for the cases of $\epsilon_{\rm WS}\le -1$ MeV.  
On the other hand, in the case of
$-1$ MeV $<\epsilon_{\rm WS}<0$ MeV 
the HF wave function 
(equivalently one quasi-particle wave function in HFB) extends largely, 
while the HFB wave function does not get much extension 
due to the pairing anti-halo effect.  

In the bottom panel of Figs. \ref{Ne}, the average pairing gaps
are shown as a function of the single particle energy $\epsilon_{\rm WS}$.  
It is seen that
the average pairing gap increases 
as $\epsilon_{\rm WS}$ approaches zero. 
This is due to the 
fact that the paring field couples with the extended wave functions 
of weakly bound nucleons in the self-consistent calculations.  
That is, the pairing field is 
extended as the wave functions do and becomes larger for a loosely bound 
system.

\begin{figure}
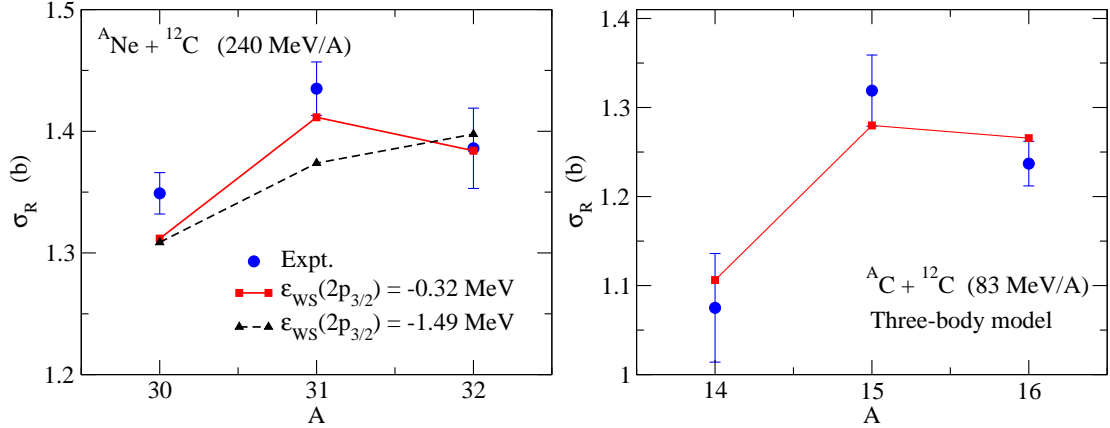
   \label{cross-fig}
\includegraphics[scale=0.45,clip]{fig3.eps}
\includegraphics[scale=0.45,clip]{fig4.eps}
\caption{(Color online) 
(left panel) Reaction cross sections of Ne isotopes on a $^{12}$C target at 
$E_{\rm lab}$=240 MeV/A.  
The cross sections are calculated with the 
Glauber theory with HF and HFB densities. 
The solid line with the filled squares shows the results 
of $S_n$($^{31}$Ne) =$|\varepsilon (2p_{3/2})|$=0.32 MeV, 
while the dashed line with 
the open triangles is obtained 
for $S_n$($^{31}$Ne) =$|\varepsilon (2p_{3/2})|$=1.49 MeV. 
The experimental data are taken from Ref. \cite{Takechi10}. (right panel)
Reaction cross sections of C isotopes on a $^{12}$C 
target at $E_{\rm lab}$=83 MeV/A.  
The cross sections are calculated with the 
Glauber theory with three-body model densities. 
The experimental data are taken from Ref. \cite{FYZ04}.
 } 
\end{figure}

Let us now calculate the reaction cross sections for the $^{30,31,32}$Ne 
isotopes and discuss the role of pairing anti-halo effect. 
To this end, we use the Glauber theory, in which we adopt the prescription 
in Refs. \cite{HSCB10,AIS00} in order to take into account 
the effect beyond the optical limit approximation. 
The left panel of Fig. 2 
shows the 
reaction cross sections of the $^{30,31,32}$Ne nuclei on a $^{12}$C target 
at 240 MeV/nucleon. 
We use the target density given in Ref. \cite{OYS92} and the profile function 
for the nucleon-nucleon scattering given in Ref. \cite{AIHKS08}. 
In order to evaluate the phase shift function, we use the two-dimensional 
Fourier transform technique \cite{BS95}. 
The cross sections $\sigma_R$ shown in Fig. 2 
 are calculated 
by using projectile densities 
constrained to two different separation energies of 
the 2$p_{3/2}$ neutron state. 
The dashed line with triangles is 
obtained using the wave functions with the separation energy 
$S_n$ ($^{31}$Ne) =$|\epsilon_{\rm WS}|$=1.49 MeV, while the solid line with squares is calculated with the 
wave functions of $S_n$($^{31}$Ne)=$|\epsilon_{\rm WS}|$=0.32 MeV. The empirical separation energy of 
$^{31}$Ne has a large ambiguity with $S_n$=0.29$\pm1.64$ MeV\cite{J07}.  
The cross section $\sigma_R$ of $^{30}$Ne is already much larger than 
the systematic values of Ne isotopes with A$<$30.  On top of that, 
we can see a clear odd-even staggering in the results with the 
smaller separation energy, $S_n$=0.32 MeV, as much as in the experimental data,  
while almost no staggering is seen in the case of the larger 
separation energy, $S_n$=1.49 MeV.  
This difference is easily understood 
by looking at the anti-halo effect for $|\epsilon_{\rm WS}|\le1$MeV shown in the 
middle panel of Fig. \ref{Ne}.  
Recently, the effect of deformation of neutron-rich Ne
isotopes on reaction cross sections was evaluated using a deformed Woods-Saxon model 
\cite{Minomo11,UHS12}.  
It was shown that the deformation is large as much as $\beta_2\sim 0.42$ 
in $^{31}$Ne 
and enhances the reaction cross section by about 5\%.  However, 
the calculated results 
did not show any significant odd-even staggering in $\sigma_R$ between $^{28}$Ne and $^{32}$Ne
 \cite{Minomo11}.  
 
We investigate next 
neutron-rich C isotopes. We particularly study the $^{16}$C nucleus 
using a three-body model given in Ref. \cite{HS07}. 
In this case, the valence neutron in $^{15}$C occupies the 2$s_{1/2}$ 
level at $\epsilon_{\rm WS}=-1.21$ MeV, while 
$^{16}$C is an admixture of mainly the $(2s_{1/2})^2$ and 
$(1d_{5/2})^2$ configurations. 
Assuming the set D given in Ref. \cite{HS07} 
for the parameters of the Woods-Saxon 
and the density distribution for $^{14}$C given in 
Ref. \cite{PZV00}, the rms radii are estimated to be 
2.53, 2.90, and 2.81 fm for $^{14}$C, $^{15}$C, and $^{16}$C, 
respectively. 
The corresponding reaction cross sections $\sigma_R$ calculated with 
the Glauber theory are shown in the right panel of Fig. 2.  
The calculation well reproduces the experimental odd-even staggering 
of the reaction cross sections, that is a clear manifestation of 
the pairing anti-halo effect.

In order to quantify the OES of reaction cross sections, we 
introduce the staggering parameter defined by 
\be
\gamma_3=(-)^{A}\frac{\sigma_R(A+1)-2\sigma_R(A)+\sigma_R(A-1)}{2},
\label{EQ-gam}
\ee
where $\sigma_R(A)$ is the reaction cross section of a nucleus
 with mass number $A$.  We can define the 
same quantity also for rms radii. 
Notice that this staggering parameter 
is similar to the one often used for the OES of binding 
energy, that is, the pairing gap.

\begin{figure} \label{stagger}
\begin{center}
\includegraphics[scale=0.6,clip]{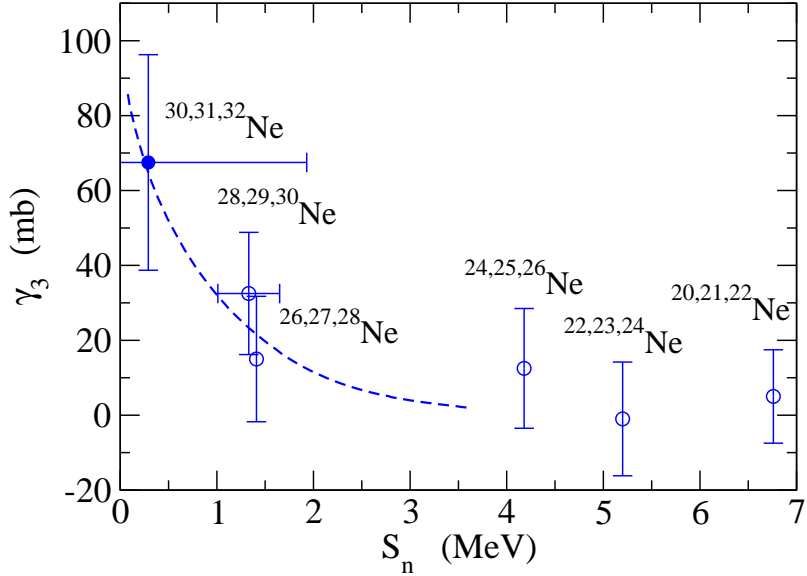}
\end{center}
\caption{(Color online) 
The staggering 
parameter $\gamma_3$ of reaction cross sections 
defined by Eq. (\ref{EQ-gam}) 
for the Ne isotopes with the $^{12}$C target at $E$=240 MeV/nucleon.  
This is plotted 
as a function of the neutron separation energy S$_n$ of the odd-$A$ nuclei.  
The experimental data 
for the reaction cross sections are taken from Ref. \cite{Takechi10}, while 
the empirical separation energies are taken from Refs. \cite{Audi,J07}. 
The dashed line is the calculated staggering parameter for 
the $^{30,31,32}$Ne isotopes, assuming that the valence neutron 
of 
$^{31}$Ne occupies the 2p$_3/2$ orbit. 
}
\end{figure}

The experimental staggering parameters $\gamma_3$ are plotted in Fig. \ref{stagger}
for Ne isotopes as a function of the neutron separation energy for the 
odd-mass nuclei. We use the experimental reaction cross sections given 
in Ref. \cite{Takechi11} while we evaluate 
the separation energies with 
the empirical binding energies listed in Ref. \cite{Audi}. 
For the neutron separation energy for the $^{31}$Ne nucleus, we use the value 
in Ref. \cite{J07}. The experimental uncertainties of the staggering parameter 
are obtained as
\begin{equation}
\delta\gamma_3=\frac{\sqrt{(\delta\sigma_R(A+1))^2+4(\delta\sigma_R(A))^2
+(\delta\sigma_R(A-1))^2}}{2},
\end{equation}
where $\delta\sigma_R(A)$ is the experimental uncertainty for the 
reaction cross section of a nucleus with mass number $A$. 
The figure also shows by the dashed line 
the calculated staggering parameter for the 
$^{30,31,32}$Ne nuclei with the 2p$_{3/2}$ orbit.
One sees that the 
experimental staggering parameter 
agrees with the calculated value for 
$^{30,31,32}$Ne nuclei when one assumes that the valence neutron in 
$^{31}$Ne occupies the 2p$_{3/2}$ orbit. 
Furthermore, although the structure of lighter odd-A Ne isotopes is not 
known well, it is interesting to see that 
the empirical staggering parameters closely follow the 
calculated values for the 2p$_{3/2}$ orbit. 
This may indicate that the low-$l$ single-particle 
orbits are appreciably mixed in 
these Ne isotopes due to the deformation effects\cite{H10,UHS11}.

\section{Isospin Dependent Pairing Interaction and  Pairing Gaps in Finite Nuclei}

There are mainly two different approaches for a 
calculation of pairing correlations in finite nuclei. 
The first approach is based on phenomenological pairing interactions
whose parameters are determined using some selected data 
\cite{dob01},
while the second approach starts from a bare
nucleon-nucleon interaction and eventually includes the effect of
phonon coupling~\cite{bar99}. The latter approach
 has shown that
the medium polarization 
reduces the pairing gaps in neutron matter while,  
in symmetric matter, the neutron pairing gaps  
are  much enlarged at low density
compared to that of the bare calculation.
This enhancement takes place especially for 
neutron Fermi momenta $k_{\mathrm{Fn}}<0.7$~fm$^{-1}$.

We propose  effective density-dependent
pairing interactions which reproduce both the neutron-neutron ($nn$) 
scattering length at zero density and the neutron pairing gap in
uniform matter.
In order to simultaneously describe the density dependence of the neutron
pairing gap for both symmetric and neutron matter,
it is  necessary to include an isospin dependence in the
effective pairing interaction \cite{mar07}.
Depending on whether the medium polarization effects on the pairing
gap given in Ref.~\cite{cao06} are taken into account or not, 
we invent  two different density dependent functionals in the pairing
interaction.  Then, we apply these interactions to study  pairing gaps in semi-magic finite 
 nuclei, such as Ca, Ni, Sn and Pb isotopic chains.

The density$-$dependent pairing interaction can be read as
\be
v_{\rm pair}(1,2)=
\frac{1-\mathrm{P}_\sigma}{2}
v_0 \,\mathrm{g}[\rho,I] 
\,\delta(\rv_1-\rv_2).
\label{IV-pair}
\ee
where $\rho$ is the nuclear density and
$I$ is defined as $I=(\rho_n -\rho_p)/\rho\cdot\tau_z$~\cite{mar07}.  
In Ref.~\cite{mar07},  an isovector
dependence in the density-dependent term $\mathrm{g}$ is separated 
  to two parts  
 $\mathrm{g}=\mathrm{g}^1+\mathrm{g}^2$. 
The function $\mathrm{g}^1$ is determined to mimic 
 the bare pairing gaps in nuclear matters and the function $\mathrm{g}^2$ 
takes care of  the medium polarization 
effect. The functional form of $\mathrm{g}^1$ is 
given by
\be
\mathrm{g}^1[\rho,I] =  1
-\mathrm{f}_\mathrm{s}(I)\eta_\mathrm{s}
\left(\frac{\rho}{\rho_0}\right)^{\alpha_\mathrm{s}}
-\mathrm{f}_\mathrm{n}(I)\eta_\mathrm{n}
\left(\frac{\rho}{\rho_0}\right)^{\alpha_\mathrm{n}} \;, 
\label{eq:g1t}
\ee
where $\rho_0$=0.16~fm$^{-3}$ is the saturation density of symmetric nuclear
matter and   the functions $\mathrm{f}_\mathrm{s}(I)$ and $\mathrm{f}_\mathrm{n}(I)$ are
$\mathrm{f}_\mathrm{s}(I)=1-\mathrm{f}_\mathrm{n}(I)$ and 
$\mathrm{f}_\mathrm{n}(I)=I=(\rho_\mathrm{n}(r)-\rho_\mathrm{p}(r))/\rho(r)\cdot\tau_z$. 
 The values of parameters $\eta_\mathrm{s}, \eta_\mathrm{n}$ and  powers of density 
dependence $\alpha_\mathrm{s}, \alpha_\mathrm{n}$ are given elsewhere~\cite{mar07}.

\begin{figure}[htb]\label{gaps}
\begin{center}
\includegraphics[width=0.7\textwidth,clip]{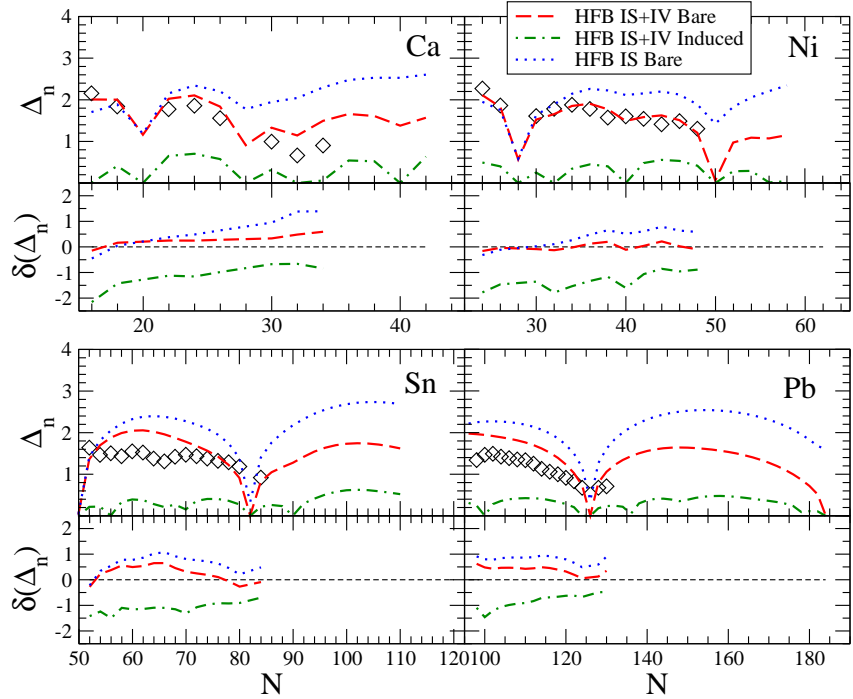}
\caption{Comparison of the neutron HFB pairing gaps 
 $\Delta_\mathrm{n}$ 
with the odd-even mass staggering
 given by the three-point formula $\Delta^{(3)}$.
The dotted line shows the results of 
the pairing interactions IS+IV~Bare, 
while the short dashed, and long dashed lines are obtained
with the pairing interactions  IS+IV~Screened  and IS~Bare,
respectively. 
The difference $\delta(\Delta_\mathrm{n})$ is defined as 
$\delta(\Delta_\mathrm{n})=\Delta_\mathrm{n}(\mathrm{th.})-\Delta_\mathrm{n}(\mathrm{exp.})$. All units are given in MeV.}
\label{fig:1}
\end{center}
\end{figure}

We perform  Hartree-Fock-Bogoliubov (HFB) calculations for 
semi-magic Calcium, Nickel, Tin and Lead isotopes 
using these density-dependent pairing interactions in Eq. (\ref{IV-pair})  
derived from a microscopic nucleon-nucleon interaction.
Our calculations reproduce well 
 the  neutron number dependence of experimental data 
 for  binding energy, 
two neutrons separation energy, and odd-even mass staggering
of these isotopes~\cite{mar08}.  
Especially the interaction IS+IV Bare without the medium polarization effect 
 gives  satisfactory 
results of all the isotopes as is seen in Fig. \ref{gaps}.  
  It is clear in  the comparison between
IS bare and IS+IV bare that the isospin dependence of the 
pairing interaction plays an important role in the pairing gaps 
  of neutron-rich
nuclei. 
The isospin dependent pairing interaction is further applied to  both 
even-even and even-odd nuclei by using EV8-odd program. 
The  results successfully reproduce  the empirical 
 isotope and isotone dependent functionals of the odd-even mass differences of
 several medium-heavy and heavy nuclei~\cite{Bert}.
Yamagami et al., studied also the quadratic isospin dependent terms of pairing interaction
\cite{Yamagami,Yama12}.  They adopted several Skyrme interactions with different effective mass and
showed the effective mass dependence of the isospin dependent terms.  It is also shown in 
 ref. \cite{YN11} that the effect of Coulomb two-body interaction on the proton pairing gaps 
 can be mimicked by the leaner isospin dependent term in the pairing interaction.

\section{Summary}
We have studied the mass radii of Ne isotopes with the Hartree-Fock (HF) and 
Hartree-Fock-Bogoliubov (HFB) methods with a Woods-Saxon potential. 
The reaction cross sections 
$\sigma_R$ 
were calculated using the Glauber 
theory with these microscopic densities. 
We have shown that the empirical odd-even staggering in the 
reaction cross sections of neutron-rich 
Ne isotopes with the mass A=$30\sim32$ is 
well described by the HFB density and can be 
considered as a clear manifestation of the pairing 
anti-halo effect associated with 
a loosely-bound 2$p_{3/2}$ 
wave function.  The index of the odd-even staggering is proposed and applied 
successfully for the study of the  reaction cross sections of Ne isotopes to revel 
 the role of the pairing correlations in exotic nuclei.  

A new type of density dependent contact pairing interaction 
was obtained to reproduce
the microscopic pairing gaps in symmetric and neutron matter.
We performed also HFB  calculations for 
semi-magic Calcium, Nickel, Tin and Lead isotopes 
using these density-dependent pairing interactions.  
Our calculations reproduce well  the  neutron number dependence of 
experimental data  of binding energy, 
two neutrons separation energy, and odd-even mass staggering
of these isotopes with  the interaction IS+IV Bare.
Recently, the isospin dependent interaction is extended to introduce the quadratic terms of 
isospin and applied to a wide region of nuclei in the mass table.
These result suggests that by introducing the isospin dependent  term 
in the pairing interaction,
one can construct a global effective pairing interaction which is 
applicable to nuclei in a wide range of the nuclear chart.

\section*{Acknowledgments}

This work was supported by the Japanese
Ministry of Education, Culture, Sports, Science and Technology
by Grant-in-Aid for Scientific Research under
the program numbers (C) 22540262.


\end{document}